\documentstyle[12pt]{article}
\normalbaselines
\newcommand{\be}{\begin{eqnarray}}
\newcommand{\ee}{\end{eqnarray}}
\begin{document}
\sl
\bibliographystyle{unsrt}

\vbox{\vspace{6mm}}

\begin{center}
{\large\bf $SU(2)$ ACTION-ANGLE VARIABLES}\\[7mm]
{\bf Demosthenes Ellinas}\\
%Department of Theoretical Physics, University of Helsinki\\
%Siltavuorenpenger 20, SF-00170 Helsinki, Finland \\
%{\it ellinas@phcu.helsinki.fi}\\
Departamento de Fisica Teorica, Faculdad de Fisica \\
Universidad de Valencia, E-46100 Burjasot Valencia Spain\\
{\it ellinas@evalvx.ific.uv.es}\\
\vspace{10mm}
{\bf Abstract}
\end{center}

%\vspace{2mm}
%\begin{abstract}
 Operator angle-action variables are studied in the frame of the $SU(2)$
algebra, and their eigenstates and coherent states are discussed. The
quantum mechanical addition of action-angle variables is shown to lead
to a novel non commutative Hopf algebra. The group contraction is used
to make the connection with the harmonic oscillator.
%\end{abstract}

\vspace{5mm}

\begin {center}
Presented at the "Harmonic Oscillators" Workshop

University of Maryland , USA ,March 1992.
\end{center}

\pagebreak

\section{Introduction}

  Action-angle variables in quantum mechanics one known to lack, in the
operator
level, some of properties of their classical analogues [1,2]. Especially
the exponential phase operators for the harmonic oscillator, occuring in
the polar decomposition of the bosonic creation and annihilation operators
(an operator analogon of the polar decomposition of a complex number),
lack the unitary and satify the weaker condition of one side-unitary or
isometry operator. Based on the mathematical fact that, unlike in finite
dimensional Hilbert spaces as the Fock space of harmonic oscillator,
in finite spaces an isometry is equivalent to a unitary operator, we have
in recent works, suggested a group theoretical construction of a unitary
phase operator by introducing action-angle variables for the $SU(2)$ algebra
and going over to their oscillator counterparts via the In\"on\"u-Wigner
method of group contraction [3-6]. In this report we will briefly review and
then expand this work with respect to two aspects: first, a set of coherent
states will be introduced along the lines of the displacement operator
creating the usual coherent states from the vacuum state and second,
we will show that addition of spins in terms of their action-angles (polar)
operators, unlike the usual addition in terms of the step (cartesian)
operators, involves a genuine no commutative, no co-commutative Hopf algebra
structure and relates interestingly the phase operators subject to the
subject of quantum groups.

\section{Action-angle Variables and States}
Let us start with the $SU(2)$ action-angle operators
\be
J_-=e^{i\Phi}\sqrt{J_+J_-}=\sqrt{J_-J_+}e^{i\Phi}
\ee
\be
J_+=e^{-i\Phi}\sqrt{J_-J_+}=\sqrt{J_+J_-}e^{-i\Phi}
\ee
where
\be
J_+=\sum\limits^{2j}_{m=0}\sqrt{m(2j-m+1)}|J;m+1><J;m|\hspace{1.0cm},
\hspace{1.0cm}J_-=J^+_+
\ee
\be
J_3=\sum\limits^{2j}_{m=0}(m-j)|J;m><J;m|
\ee
and
\be
e^{i\Phi}=\sum\limits^{2j}_{\ell=0}|J;\ell><J;\ell+1|\ ,
\ee
mod$(2j+1)$, and $hh^+=h^+h={\bf 1}$ with $h\equiv e^{i\Phi},\ h^+\equiv
e^{-i\Phi}$ the unitary angle operator. Then from the fact that $h$,
generates the cyclic group $Z_{2j+1}$ acting as a cyclic permutation
in the weight space of the algebra we can construct phase states
\be
|\Phi;k>=F|J;k>=\frac{1}{\sqrt{2j+1}}\sum\limits^{2j}_{m=0}\omega^{kn}|
J;n>
\ee
through the finite Fourier transform $FF^+=F^+F=1$, which maps action
eigenstates to angle eigenstates and conjugates the respective variables, where
$\omega=\exp i(2\pi/2j+1)$. Indeed, if $g:=\omega^{J_3+j{\bf 1}}$ then
$FgF^+=h,\ FhF^+=g^{-1}$ and $g(h)$ acts as step operator in the angle
(action) state basis, i.e,
\be
h|J;n>=|J;n+1>\ \ ,\ \ \ h|\Phi;m>=\omega^m|\Phi;m>
\ee
while
\be
g^{-1}|\Phi;n>=|\Phi;n+1>\ \ ,\ \ \ g|J;m>=\omega^m|J;m>
\ee
mod$(2j+1)$ and $h^{2j+1}=g^{2j+1}={\bf 1},$ (notice that the state $|J;n>$ and
$|\Phi;m>$ where denoted as $|n>$ and $|\varphi_m>$ respectively, in Refs.
3-6). The noncommutativity between the action and the angle variables is
best expressed by the formula
\be
\omega gh=hg
\ee
which resembles the exponential form of the Heisenberg canonical commutation
relations (CR) as were originally written by Weyl with the association that
here the action operator $J_3$ is a finite version of the position operator
and the angle operator stands for the momentum operator. By virtue of
this analogy we may interpret eqs. (7-8) as the translations along the two
different directions of the phase space of our problem, which due to the
module condition is a lattice torus, parametrized by the discrete action
and angle values. Also eq. (9), exhibits the unusual noncommutative
character of two succesive translations along different directions.
Moreover, the effect of group contraction which is discussed below, is to
increase the density of the lattice points until the continous limit
$j\rightarrow\infty$. Furthermore this association to position and momentum
suggests that we should look for the "number states" $|N;m>,\ m=0,1,...,
2j$ in our finite system. Indeed by diagonalizing the finite Fourier
transform $F|N;m>=i^m|N;m>,$ we find the number states $|N;m>$, related e.g.
with the orthonormal action states as:
\be
|N;k>=\sum\limits^{2j}_{m=0}|J;m><J;m|N;k>\ ,
\ee
with expansion coefficients given in terms of the Hermite polynomial,
$H_k$ with discrete argument,
\be
<J;m|N;k>=\sum\limits^\infty_{p=-\infty}e^{-\frac{\pi}{2j+1}(p(2j+1)+m)^2}
H_k\biggl( \sqrt{\frac{2\pi}{2j+1}}(p(2j+1)+m)\biggr)
\ee
This situation is akin to that of the harmonic oscillator number states which
are similarly eigenstates of the usual Fourier transform operator which
conjugates position and momentum operators, a fact that stems from the
property of the oscillator eigenstates $\exp(-\frac{1}{2}^{\times 2}) H_k(x)$,
to be their own Fourier transforms. Especially the vacum or lowest
number state is,
\be
|N;0>=\sum\limits^{2j}_{m=0}\omega^{\frac{i}{2}m^2} \theta_3(im|i(2j+1))
|J;m>
\ee
where $\theta_3$ is the theta-Jacobi function [7]:
\be
\theta_3(z|\tau)=\sum\limits^\infty_{s=-\infty}e^{zi2\pi s+\tau i\pi s^2}\ .
\ee
Having the action $|J;m>$, the angle $|\Phi;n>$ and the number states
$|N;k>$ as were given above, we can further built, as have been outlined in
Ref. 4, the quantum theory of action-angle variables by
introducing the corresponding coherent states acting on the vacum
$|N;0>$, with a displacement operator. Such an operator  is furnished
by the unitary traceless elements $J_{m_1,m_2}:=\omega^{m_1m_2/2}g^{m_1}
h^{m_2}$, where $J^+_{m_1,m_2}=J_{-m_1,-m_2}=J_{2j+1-m_1,2j+1-m_2}$,
with $(m_1,m_2)$ pairs belonging to the square index-lattice $0\leq
m_1,m_2\leq 2j$ with boundary conditions and the $(0,0)$ pair excluded.

The following interesting properties of these operators suggest them as the
Glauber displacement operator of our case; first they constitute
an orthonormal set of $(2j+1)^2-1$ elements obeying the relation
\be
<J_{\vec{m}},J_{\vec{n}}>:=Tr\ J_{\vec{m}}J_{\vec{n}}=(2j+1)\delta_{\vec{m}
+\vec{n},\vec{0}}\ ,
\ee
where e.g. $J_{\vec{m}}=J_{m_1m_2}$, and further,
\be
J_{\vec{m}}J_{\vec{n}}=\omega^{-\frac{1}{2}\vec{m}\times\vec{n}}J_{\vec{m}
+\vec{n}}
\ee
and
\be
J_{\vec{n}}J_{\vec{m}}=\omega^{\vec{m}\times\vec{n}}J_{\vec{m}}J_{\vec{n}}
\ee
and finally
\be
[J_{\vec{m}},J_{\vec{n}}]=-2i\sin\biggl[\frac{\pi}{2j+1}\vec{m}\times
\vec{n}\biggr]J_{\vec{m}+\vec{n}}
\ee
mod$(2j+1)$, while $\vec{m}\times\vec{n}=m_1n_2-m_2n_1$.
With the aid of these operators we now introduce coherent states
$|\vec{\ell}>$, for the action-angle system by acting on the vacum:
\be
|\vec{\ell}>:=J_{\vec{\ell}}|N;0>=\omega^{\frac{3}{2}\ell_1\ell_2}
\sum\limits^{2j}_{m=0}\omega^{\ell_1m+\frac{i}{2}m^2}\theta_3(im|i(2j+1))
|J;m+\ell_2>
\ee
These are now coherent states defined on the lattice phase space which
is the appropriate phase space of the quantum action-angle variables.
They involve the Jacobi theta functions which are also appearing in the case
of the ordinary coherent states when, looking for a complete subset out
of the over complete set of coherent states we lattice the phase space.
Elsewhere, the normalization and minimum uncertainty properties of the states
will be studied in detail.

\section{Quantum Angles Addition}
Let us now turn to the case where there are several action-angle degrees of
free
dom
and search for the way we combine them quantum mechanically. The similar
problem for the "cartesian" generators $J_i$, with $[J_i,J_j]=2i\epsilon_{ijk}
J_k$
is the fundamental theme of addition of spins and customanily is solved by
tensoring the generators,
\be
\Delta  J_i:=J_i\otimes{\bf 1}+{\bf 1}\otimes J_i
\ee
which again satisfy the commutation relations, $[\Delta J_i,\Delta J_j]=
2i\epsilon_{ijk}\Delta J_k$. In our case, for the "polar" generators
$g=\omega^{(J_3+j{\bf 1})}$ and $h=\omega^{F(J_3+j{\bf 1})F^+}$ with
$\omega gh=hg$ we must find an appropriate tensoring (coproduct in the
jargon of Hopf algebras), which provides such $\Delta g$ and $\Delta h$
that $\omega\Delta g=\Delta h$. Two such coproducts we have found,
\be
\Delta g=g\otimes g\ \ \ ,\ \ \ \ \Delta h=h\otimes{\bf 1}+g\otimes h
\ee
and
\be
\Delta g=g\otimes g\ \ \ ,\ \ \ \ \Delta h=h\otimes g+g^{-1}\otimes h
\ee
which both have the remarkable property of not been the same under permutation
of their components involved in the tensor products. This is  distinctly
differe
nt
to the usual addition of spins, where there is no sence of order in the
tensoring the spins. Technically speaking we have here a  natural case of
no co-commutativity unlike in eq. (19), where the product is co-commutative
[8-11]. We end here this discussion, as we intent to expand it elsewhere,
by saying that it is also possible to show the Hopf and quasi triangular
Hopf algebra structure of the above tensoring and then to find the R-matrix and
to verify the Yang-Baxter equation.

\section{Contraction to the Oscillator}
Before we came to conclusions let us mention that as was shown in Ref. 3
via the group contraction that the $SU(2)$ action-angle variables can be
contructed to those of the oscillator and the dynamical aspects of this
procces could be exemplified by studing the Jaynes-Cummings model. We
illustrate now this idea be contracting the $SU(2)$ generators to
the oscillator generators in the Bargmann analytic realization. In the
space of analytic polynomials of degree $2j$ the $SU(2)$ algebra is
realized as,
\be
J_+=-z^2\frac{d}{dz}+z2j\hspace{1.5cm}
J_-=\frac{d}{dz}\hspace{1.5cm}
J_3=z\frac{d}{dz}-j
\ee
where $z$ is the complex label of the spin coherent states, and geometrically
stands for the projective coordinate of the coset sphere $SU(2)/U(1)\sim
S^1$. Transforming now the generators like $J_\pm \rightarrow J_\pm/\sqrt{2j}$
and $J_3\rightarrow J_3+j{\bf 1}$ we find in the large $j$ limit, the
oscillator generators in their Bargmann form as follows:
\be
\frac{J_+}{\sqrt{2j}}=-\frac{(\sqrt{2j}z)^2}{2j}\frac{d}{d(\sqrt{2j}z)}
+\sqrt{2j}z\approx\alpha=a^+
\ee
\be
\frac{J_-}{\sqrt{2j}}=\frac{d}{d(\sqrt{2j}z)}\approx\frac{d}{d\alpha}=a
\ee
and
\be
J_3+j=\sqrt{2j}z\frac{d}{d(\sqrt{2j}z)}\approx\alpha\frac{d}{d\alpha}=N
\ee
where $\sqrt{2j}z\approx\alpha$ is the complex variable of the Glauber coherent
states which is now becoming the coordinate of the tangent phase plane of the
harmonic oscillator. One can further show that the overlap, the completeness
relation and all other notions of the spin coherent states can be contracted
to their respective oscillator counterparts. Moreover in Ref. 5 has been
shown how a $q$-deformed oscillator with $q$ deformation parameter to be
root of unity can be employed to define action-angles variables in a
finite Fock Hilbert space and a number of their properties have been
worked out. In such an approach we have shown [5], that the contraction
method is substituted by the limit procedure of undeforming the
$q$-oscillator to the usual ocillators.

\section{Conclusion}
In conclusion, we have shown that the quantization of action-angle classical
variables can be developed in the framework of the $SU(2)$ algebra in a
manner which allows for the classical properties of these variables to
find well defined operator analogues. Interesting relations to the quantum
groups and Hopf algebras are naturally emerge from the present method
of angle quantization which will be pursued further, together with the
introduction of the Wigner function for the action-angles variables and
the star and Moyal product defined between functions of the phase space of
our problem.
\vskip 2.0cm

I wish to thank the DIGCYT (Spain) for finacial support.

\end{document}